\documentstyle[12pt,epsfig]{article}
\newcommand{\nc}{\newcommand}
\nc{\fdiag}{0}
\nc{\bg}{B. Grzadkowski}
\nc{\BG}{Bohdan Grzadkowski}
\nc{\lsp}{\;\;\;\;\;\;\;\;}
\nc{\beq}{\begin{equation}}   \nc{\eeq}{\end{equation}}
\nc{\bea}{\begin{eqnarray}}   \nc{\eea}{\end{eqnarray}}
\nc{\baa}{\begin{array}}      \nc{\eaa}{\end{array}}
\nc{\bit}{\begin{itemize}}    \nc{\eit}{\end{itemize}}
\nc{\ben}{\begin{enumerate}}  \nc{\een}{\end{enumerate}}
\nc{\bce}{\begin{center}}     \nc{\ece}{\end{center}}
\nc{\non}{\nonumber}
\nc{\lumun}{\;{\hbox {fb}^{-1}}{\hbox {yr}^{-1}}}
\nc{\hc}{\hbox {h.c.}}
\nc{\re}{\hbox {Re}}
\nc{\im}{\hbox {Im}}
\nc{\etal}{\hbox{et al.}}
\nc{\pbarn}{\;\hbox {pb}}
\nc{\prd}[3]{{\it Phys.\ Rev.}\ {{\bf D{#1}} (#2) #3}}
\nc{\prl}[3]{{\it Phys.\ Rev.\ Lett.}\ {{\bf {#1}} (#2) #3}}
\nc{\plb}[3]{{\it Phys.\ Lett.}\ {{\bf B{#1}} (#2) #3}}
\nc{\npb}[3]{{\it Nucl.\ Phys.}\ {{\bf B{#1}} (#2) #3}}
\nc{\ptp}[3]{{\it Prog.\ Theor.\ Phys.}\ {{\bf {#1}} (#2) #3}}
\nc{\zfp}[3]{{\it Z.\ Phys.}\ {{\bf C{#1}} (#2) #3}}
\nc{\mpla}[3]{{\it Mod.\ Phys.\ Lett.}\ {{\bf A{#1}} (#2) #3}}
\nc{\rmp}[3]{{\it Rev.\ Mod.\ Phys.}\ {{\bf {#1}} (#2) #3}}
\nc{\ijmpa}[3]{{\it Int.\ J.\ of\ Mod.\ Phys.}\
               {{\bf A{#1}} (#2) #3}}
\nc{\app}[3]{{\it Acta\ Phys.\ Polon.}\ {{\bf B{#1}} (#2) #3}}
\nc{\epj}[3]{{\it Eur. Phys. J.}\ {{\bf C{#1}} (#2) #3}}
\nc{\prep}[3]{{\it Phys.\ Rep.}\ {{\bf {#1}} (#2) #3}}
\nc{\ra} {\rightarrow}
\nc{\cw}{\cos\theta_W}        
\nc{\sw}{\sin\theta_W}
\nc{\ttbar}{t\bar{t}}
\nc{\bbbar}{b\bar{b}}
\nc{\tanb} {\tan \beta}
\nc{\cotb} {\cot\beta}
\nc{\twbdec} {t\rightarrow W^+ b}
\nc{\tbwbdec} {\bar{t} \rightarrow W^- \bar{b}}
\nc{\hprod} {e^+e^- \ra Z^\ast \ra H Z}
\nc{\epem} {e^+e^-}
\nc{\wpwm} {W^+W^-}
\nc{\tbar} {\bar{t}}
\nc{\bbar} {\bar{b}}
\nc{\wpp} {W^+}
\nc{\mt}{m_t}
\nc{\mts}{m_t^2}
\nc{\mw} {m_W}
\nc{\mws} {m_W^2}
\nc{\mz} {m_Z}
\nc{\mzs} {m_Z^2}
\nc{\mh} {m_H}
\nc{\mhs} {m_H^2}
\nc{\ma} {m_A}
\nc{\mas} {m_A^2}
\nc{\hdec}{H \ra t\bar{t}}
\nc{\ttbardec}{\ttbar \ra W^+W^-\bbbar}
\nc{\po}{\Phi_1}
\nc{\pod}{\Phi_1^\dagger}
\nc{\pht}{\Phi_2}
\nc{\phtd}{\Phi_2^\dagger}
\nc{\phtt}{{\tilde{\Phi}}_2}
\nc{\popo}{\po^\dagger\po}
\nc{\phtpt}{\pht^\dagger\pht}
\nc{\popt}{\po^\dagger\pht}
\nc{\phtpo}{\pht^\dagger\po}
\nc{\sq}{\sqrt{2}}
\nc{\nsd} {N_{SD}}
\nc{\ntt} {N_{tt}}
\nc{\vs}{\vspace{2mm}}

\nc{\sty}{\hat{S}^t_1} \nc{\pty}{\hat{P}^t_1}
\nc{\sts}{(\sty)^2}      \nc{\pts}{(\pty)^2}
\nc{\yts}{\sts+\pts}
\nc{\sby}{\hat{S}^b_1} \nc{\pby}{\hat{P}^b_1}
\nc{\sbs}{(\sby)^2}      \nc{\pbs}{(\pby)^2}
\nc{\ybs}{\sbs+\pbs}

\nc{\styi}{\hat{S}^t_i} \nc{\ptyi}{\hat{P}^t_i}
\nc{\stsi}{(\styi)^2}      \nc{\ptsi}{(\ptyi)^2}
\nc{\ytsi}{\stsi+\ptsi}
\nc{\sbyi}{\hat{S}^b_i} \nc{\pbyi}{\hat{P}^b_i}
\nc{\sbsi}{(\sbyi)^2}      \nc{\pbsi}{(\pbyi)^2}
\nc{\ybsi}{\sbsi+\pbsi}

\def\vev#1{\langle #1 \rangle}
\def\ie{{\it i.e.}}

\def\sb{s_\beta}
\def\cb{c_\beta}
\def\rts{\sqrt s}

\def\hsm{h_{\rm SM}}
\def\mhsm{m_{\hsm}}

\def\h{h}

\def\hl{h^0}
\def\hh{H^0}
\def\ha{A^0}

\def\hpm{H^{\pm}}
\def\mhl{m_{\hl}}
\def\mhh{m_{\hh}}

\def\lsim{\mathrel{\raise.3ex\hbox{$<$\kern-.75em\lower1ex\hbox{$\sim$}}}}
\def\gsim{\mathrel{\raise.3ex\hbox{$>$\kern-.75em\lower1ex\hbox{$\sim$}}}}
\def\anti{\overline}

\def\fbi{~{\rm fb}^{-1}}
\def\fb{~{\rm fb}}

\def\gev{\,{\rm GeV}}

\begin{document}
%
\font\fortssbx=cmssbx10 scaled \magstep2
\medskip
\begin{flushright}
$\vcenter{
\hbox{\bf UCD-2000-01} 
\hbox{\bf IFT/99-33}
\hbox{\bf LC-TH-2000-022}
\hbox{\bf hep-ph/0001093}
\hbox{December, 1999}
}$
\end{flushright}
\vspace*{2cm}
\begin{center}
{\large{\bf Search Strategies for Non-Standard Higgs Bosons at Future 
{\boldmath $\epem$} Colliders}}\\ 
\rm
\vspace*{1cm}
\renewcommand{\thefootnote}{\alph{footnote})}
{\bf \BG$^1$}
\footnote{E-mail:{\tt bohdang@fuw.edu.pl}} 
{\bf John F. Gunion$^2$}
\footnote{E-mail:{\tt jfgucd@higgs.ucdavis.edu}} 
and {\bf Jan Kalinowski$^1$}
\footnote{E-mail:{\tt kalino@fuw.edu.pl}}\\

\vspace*{1.5cm}
{$^1$ \em Instytut Fizyki Teoretycznej UW, Hoza 69, 
Warsaw, Poland}\\
{$^2$ \em Davis Institute for High Energy Physics, 
UC Davis, CA, USA }\\

\vspace*{1.5cm}

{\bf Abstract}

\end{center}
\vspace{5mm}    Already in the simplest
two-Higgs-doublet  model with CP violation in the
Higgs sector, the $3\times3$ mixing matrix for
the neutral Higgs bosons can substantially
modify their couplings, thereby endangering the
``classical'' Higgs search strategies.  However,  there are sum rules
relating Yukawa and Higgs--$Z$ couplings which ensure 
that the $ZZ$, $b\anti b$ and $t\anti t$ couplings of
a given neutral 2HDM Higgs boson cannot all be simultaneously suppressed.
This result implies that any single Higgs boson 
will be detectable at an $e^+e^-$ collider if the
$Z$+Higgs, $b\anti b+$Higgs {\it and} $t\anti t+$Higgs production
channels are all kinematically accessible {\it and} if 
the integrated luminosity is sufficient.  We explore, as a function
of Higgs mass, the luminosity
required to guarantee Higgs boson detection, and find that
for moderate $\tanb$ values the needed luminosity 
is unlikely to be available for all possible mixing scenarios.
Implications of the sum rules for Higgs
discovery at the Tevatron and LHC are briefly discussed.

\vfill
\setcounter{page}{0}
\thispagestyle{empty}
\newpage

\renewcommand{\thefootnote}{\sharp\arabic{footnote}}
\setcounter{footnote}{0}

\section{Introduction}

Spontaneous gauge symmetry breaking in the
Standard Model (SM) is realized by introducing a single CP-even Higgs
boson, $\hsm$. The ``standard'' Higgs hunting strategies at an $\epem$
collider rely on the Higgs-strahlung process, $\epem\to Z\hsm$, and
(for higher energies and heavier Higgs bosons) on the $WW$ fusion
process, $\epem\to \nu \bar{\nu} \hsm$ ($ZZ$ fusion is smaller by an order
of magnitude) \cite{lcrep}. However, 
even the simplest two-Higgs-doublet
model (2HDM) extension of the SM exhibits a rich Higgs sector structure.
Moreover, it allows for spontaneous and/or explicit CP
violation in the scalar sector~\cite{weinberg}. CP violation, which in 
the SM is achieved only  by the Yukawa couplings of the Higgs boson to quarks
being explicitly complex \cite{km}, 
could equally well be partially or wholly due to new physics beyond
the SM.  The possibility that an extended Higgs sector is
responsible for CP violation is particularly appealing,
especially as a means for obtaining an adequate level
of baryogenesis in the early universe \cite{gavela}.

The CP-conserving (CPC) 2HDM predicts~\footnote{ The same menagerie of
  pure-CP Higgs bosons is found at the tree level in the minimal
  supersymmetric model (MSSM) \cite{hhg}. However, with $CP$-violating
  phases of soft-supersymmetry breaking terms, the $\hl$, $\hh$ and
  $\ha$ will mix beyond the Born approximation~\cite{cp-phases}.}  the
existence of two neutral CP-even Higgs bosons ($\hl$ and $\hh$, with
$\mhl\leq\mhh$ by convention), one neutral CP-odd Higgs ($\ha$) 
and a charged Higgs
pair ($\hpm$).  The situation is more complex in the 2HDM with
CP-violation (CPV) in the scalar sector. There, 
the physical mass eigenstates, $\h_i$($i=1,2,3$), are 
mixtures (specified by three mixing angles, $\alpha_i$, $i=1,2,3$) 
of the real and imaginary components of the original neutral
Higgs doublet fields; as a result, the $h_i$ have undefined CP properties.

The absence of any $\epem\to Z \hsm$ signal in LEP2 data translates
into a lower limit on $\mhsm$: the latest analysis of four LEP
experiments at $\sqrt{s}$ up to 196 GeV implies $\mhsm$ greater than
102.6 GeV \cite{newdata}. More generally, in $\epem$ collisions,
if $\mhsm<\rts-\mz$
the $\hsm$ will be discovered, assuming sufficient integrated
luminosity.  In this note, we wish to address the extent to
which the neutral Higgs bosons of an extended Higgs sector
are guaranteed to be discovered if they are sufficiently light.
The possibility of such a guarantee rests on considering not
only $Z$+Higgs production but also Higgs pair production,
$b\anti b$+Higgs production and $t\anti t$+Higgs production
and on the existence of sum rules for the Higgs boson couplings
controlling the rates for these processes.
Our analysis will be performed for a type-II 2HDM,
wherein the neutral component of one of the Higgs 
doublet fields couples only to down-type quarks and leptons
and the neutral component of the other couples only to up-type quarks.

We first remind the reader of the 2HDM (CPV or CPC) result \cite{gghk,ggk}
that if there are two light Higgs bosons, $h_1$ and $h_2$,  
then at least one will be observable in $Zh_1$ or $Zh_2$ production or 
both in $h_1h_2$ pair production. This is because of the sum rule 
\cite{pomarol,gghk} for the Higgs boson couplings 
$C_1^2+C_2^2+C_{12}^2=1$, where
$g_{ZZ\h_i} \equiv \frac{g m_Z}{c_W} C_i$ and  
$g_{Zh_ih_j} \equiv \frac{g}{2c_W} C_{ij}$
[$c_W=\cos\theta_W$, $g$ is the
SU(2) gauge coupling constant].
If none of the three processes are observed, we know that at least
one of the two Higgs masses must lie beyond the kinematic limits defined
by $\rts<\mz+m_{h_1},\mz+m_{h_2},m_{h_1}+m_{h_2}$. 
A recent analysis of LEP data
shows that the 95\% confidence level exclusion region in the
$(m_{h_1},m_{h_2})$ plane that results from the sum rule is
quite significant \cite{cpvdelphi}.

Here, we focus on the question of whether a
{\it single} neutral Higgs boson will be observed in $\epem$ collisions
if it is sufficiently light, regardless of the masses and couplings
of the other Higgs bosons.  In general, 
such a guarantee cannot be established if only the
Higgs-strahlung and Higgs-pair production processes are considered.
First, there is a
``nightmare'' scenario in which Higgs-strahlung  is inadequate
for detection of the lightest Higgs boson $h_1$ while 
all other Higgs bosons are too heavy to be kinematically
accessible. This is easily arranged by choosing model
parameters such that the $ZZh_1$ coupling is too weak
for its detection in Higgs-strahlung production while maintaining consistency
with precision electroweak constraints~\cite{mk} despite the other
Higgs bosons being heavy.
Of course, if we were to demand that the 2HDM remains perturbative up to energy
scales of order $10^{16}\gev$, then the sum rule of Ref.~\cite{espgun}
guarantees that $\sum_i C_i^2 m_{h_i^0}^2\lsim m_B^2$, where
$m_B\sim 200\gev$, in which case this scenario could not be realized
assuming that $\rts$ is substantially larger than $m_B$.
Second, it could happen that there are two light Higgs bosons, $h_1$ and
$h_2$, but one of them, e.g. the $h_2$,
has full strength $ZZh_2$ coupling. Then,
the sum rule $C_1^2+C_2^2+C_{12}^2=1$
implies that the $ZZh_1$ and $Zh_1h_2$ couplings must both be
zero at tree-level. Consequently, the
$h_2$ will be seen in $\epem\to Z^*\to Zh_2$ 
production but the $h_1$ will not be discovered in $Zh_1$ or $h_1h_2$
production, even when these processes are kinematically accessible.
Note that this scenario is completely consistent with the above-noted
GUT-scale-perturbativity sum rule.
We stress that the above cases can arise regardless of the mixing
structure, CPC or CPV, of the neutral Higgs boson sector.
  
In \cite{ggk} we derived new sum rules relating
the Yukawa $g_{f\bar{f}h_i}$ and Higgs--$Z$ couplings of the 2HDM [see
Eq.~(\ref{newsr})] which guarantee that any $h_i$ that has suppressed
$ZZh_i$ coupling must have substantial $t\anti t h_i$ and/or $b\anti b h_i$
coupling. This result implies that if the $h_i$ 
is sufficiently light for $t\anti t h_i$ to be kinematically allowed
and if the luminosity is sufficiently large, then 
the $h_i$ will be observed in at least one of the Yukawa processes
$\epem\to f\bar{f}h_i$ ($f=t$, $b$ and possibly $\tau$),
dominated by Higgs radiation from the final
state fermions .  Therefore, the
complete Higgs hunting strategy at $e^+e^-$ colliders, and at hadron
colliders as well, must include not only the Higgs-strahlung process
and Higgs-pair production but also the Yukawa processes.~\footnote{In
the context of a CP conserving 2HDM, the
relevance of the Yukawa processes when $\tanb$ is large
has been stressed already several times \cite{oldyukawa,dkz,dr}.}
However, our earlier work left open a detailed analysis of
just how much integrated luminosity was required.

In this paper, we consider
in more detail the 2HDM in the context of future $e^+e^-$
linear colliders ($\rts\sim 500-800\gev$) with integrated
luminosity $L\sim 500-1000$ fb$^{-1}$, as planned in
one-to-two years of running at TESLA. Focusing
on the case of a light Higgs boson that cannot be observed
in Higgs-strahlung or Higgs-pair production,
we determine the $L$ required so that either 
$b\anti b h_1$ or $t\anti t h_1$ production will allow $h_1$ detection.
For the worst choices of Higgs mixing angles $\alpha_i$, 
the required $L$ is quite large. 

The outline of the paper is as follows. In the next section, we discuss
the sum rules for Higgs boson couplings in the CPV 2HDM. Then, we
present numerical results for $Zh_1h_2$, $b\bar{b}h_1$ and
$t\bar{t}h_1$ cross sections at $\epem$ linear colliders running with
$\sqrt{s}=500$ and 800 GeV and address the question of measurability 
of Yukawa couplings. In the next section, 
we determine the portions of parameter
space such that unrealistically large
integrated luminosity could be required for discovery of a light $h_1$.
In the conclusions, we summarize the main points of the paper and
briefly discuss implications of the sum rules for Higgs searches at
hadronic accelerators.

\section{Higgs boson couplings and sum rules}

In the type-II two-Higgs-doublet model,
the neutral component of the
$\Phi_1$ doublet field couples only to down-type quarks and leptons
and the neutral component of $\Phi_2$ couples only to up-type quarks.
As usual, we define $\tanb\equiv v_2/v_1$, 
where $|\vev{\Phi_{1,2}^0}|=v_{1,2}/\sqrt 2$. 
As a result of the mixing (for details see \cite{ggk})
between real and imaginary parts of neutral
Higgs fields, the Yukawa interactions of the $h_i$ mass-eigenstates
are not invariant under CP. They are given by:
\begin{equation} 
{\cal L}=h_i\bar{f}(S^f_i+iP^f_i\gamma_5)f \label{coupl} 
\end{equation}
where the scalar ($S^f_i$) and pseudoscalar ($P^f_i$) couplings are
functions of the mixing angles. For up-type and down-type quarks we have 
\begin{eqnarray}
S^u_i=-\frac{m_u}{v s_\beta}R_{i2},\;\;\;\;\;
P^u_i=-\frac{m_u}{v s_\beta}c_\beta R_{i3}, 
\label{absu} \\
S^d_i=-\frac{m_d}{v c_\beta}R_{i1},\;\;\;\;\;
P^d_i=-\frac{m_d}{v c_\beta}s_\beta R_{i3}\,,
\label{absd}
\end{eqnarray}
and similarly for charged leptons.~\footnote{ 
$s_\beta=\sin\beta$, $c_\beta=\cos\beta$, and in  our 
normalization $v\equiv \sqrt{v_1^2+v_2^2}=2m_W/g=246\,\mbox{GeV}$.}
For the 2HDM, the $R_{ij}$ are elements of the orthogonal rotation matrix 
\begin{eqnarray} 
h=R\varphi=\left(\baa{ccc}
  c_1     &  -s_1c_2          &     s_1s_2  \\
  s_1c_3  & c_1c_2c_3-s_2s_3  &  -c_1s_2c_3-c_2s_3\\
  s_1s_3 & c_1c_2s_3+s_2c_3 & -c_1s_2s_3+c_2c_3 \eaa\right) 
\left(\baa{c} \varphi_1 \\ \varphi_2 \\ \varphi_3 \eaa \right),
\label{mixing}
\end{eqnarray}
[$s_i\equiv\sin\alpha_i$ and $c_i\equiv\cos\alpha_i$] which relates
the original neutral degrees of freedom~\footnote{The remaining
  degree of freedom, $\sqrt2(\cb\mbox{Im}\phi_1^0+
  \sb\mbox{Im}\phi_2^0)$, becomes a would-be Goldstone boson which is
  absorbed in giving mass to the $Z$ gauge boson.}
\begin{equation}
(\varphi_1,\varphi_2,\varphi_3)\equiv
\sqrt 2(\mbox{Re}\phi_1^0, \, \mbox{Re}\phi_2^0, \,
 s_\beta\mbox{Im}\phi_1^0-c_\beta\mbox{Im}\phi_2^0) 
\label{degrees}
\end{equation} 
of the two Higgs doublets $\Phi_1=(\phi_1^+,\phi_1^0)$ and
$\Phi_2=(\phi_2^+,\phi_2^0)$ to the physical mass eigenstates $h_i$
($i=1,2,3$).   
Without loss of generality, we assume $m_{h_1}\le
m_{h_2} \le m_{h_3}$. 

Using the above notation, the couplings of neutral Higgs and $Z$ bosons 
are given by
\begin{eqnarray}
 C_i &=& s_{\beta} R_{i2}+c_{\beta}R_{i1}
\label{zzhcoup}
\\
 C_{ij} &=& w_i R_{j3}-w_j R_{i3}
\label{zhhcoup}
\end{eqnarray}
where $w_i=s_{\beta}R_{i1}-c_{\beta}R_{i2}$.  

The conventional 
CP-conserving limit can be obtained as a special case: $\alpha_2=
\alpha_3=0$. Then, if we take $\alpha_1=\pi/2-\alpha$, $\alpha$ is the
conventional mixing angle that diagonalizes the mass-squared matrix
for $\sqrt 2\mbox{Re}\phi_1^0$ and $\sqrt 2\mbox{Re}\phi_2^0$.  The
resulting mass eigenstates are $h_1=-\hl$ $h_2=\hh$ and $\sqrt
2(s_\beta\mbox{Im}\phi_1^0-c_\beta\mbox{Im}\phi_2^0)=-\ha$, where
$\hl$, $\hh$ ($\ha$) are the CP-even (CP-odd) Higgs bosons defined
earlier for the CPC 2HDM. Of course, there are other CP-conserving
limits. For instance, by choosing $\alpha_1=\alpha_2=\pi/2$, $h_1$
becomes pure $\varphi_3=-\ha$, while it is $h_2$ and $h_3$
that are CP-even.

The crucial sum rules that potentially guarantee discovery (assuming
sufficient luminosity) of any neutral
Higgs boson that is  light enough to be kinematically
accessible in Higgs-strahlung {\it and} $b\anti b$+Higgs 
{\it and} $t\anti t$+Higgs are an automatic result of
the orthogonality of the $R$ matrix.
These sum rules \cite{ggk} involve a combination
of the Yukawa and $ZZ$ couplings of any one Higgs
  boson and require that at least one of these
  couplings has to be sizable. In particular, if $C_i\to 0$ 
(the focus of our paper) then
  orthogonality of $R$ yields
\begin{equation}
  \label{newsr}
(\hat{S}^t_i)^2 + (\hat{P}^t_i)^2 
=\left(\frac{\cos\beta}{\sin\beta}\right)^2 \,,\quad
(\hat{S}^b_i)^2 + (\hat{P}^b_i)^2 
=\left(\frac{\sin\beta}{\cos\beta}\right)^2 \,
\end{equation}
where for convenience we introduce  rescaled couplings
\begin{eqnarray}
    \label{rescal}
\hat{S}^f_i\equiv \frac{S^f_i v}{m_f}\,,~~~~~~~
\hat{P}^f_i\equiv \frac{P^f_i v}{m_f}\,,
\end{eqnarray}
$f=t,b$.~\footnote{For obvious reasons we consider the third
  generation of quarks. Similar expressions hold for for lighter
  generations.}
Eq.~(\ref{newsr}) implies
that either the $t\anti t$ or the $b\anti b$ coupling of $h_i$
must be large in the $C_i\to 0$ limit; both cannot be small.
Even in the other extreme
of $C_i\to \pm 1$, \ie\ ffull strength $ZZh_i$ coupling, 
one finds that $(\hat S_i)^2+(\hat P_i)^2\to1 $, 
for both the top and the bottom quark
couplings, in the limit of either very large or very small $\tanb$.
A completely general result following from orthogonality of $R$, 
that is independent of $C_i$, is
\begin{eqnarray}
  \label{finalsr}
  \sin^2\beta [(\hat{S}^t_i)^2 + (\hat{P}^t_i)^2]
+ \cos^2\beta [(\hat{S}^b_i)^2 + (\hat{P}^b_i)^2]=1\,,  
\end{eqnarray}
again implying that the
Yukawa couplings to top and bottom quarks cannot be simultaneously
suppressed.  As a result, if an $h_i$ is sufficiently light, its
detection in association with $b\anti b$ or $t\anti t$ 
should, in principle, be possible, irrespective of the neutral Higgs sector
mixing and regardless of whether or not it is seen in $\epem\to Zh_i$
or Higgs pair production. However, this leaves open the question
of just how much luminosity is required to guarantee detection.

\section{ Higgs boson production in \boldmath $e^+e^-$ colliders}

To treat the three processes: 
(i) bremsstrahlung off the $Z$ boson ($e^+e^-\ra Zh_i$), (ii) Higgs pair
production ($e^+e^-\ra h_ih_j$), and (iii) the Yukawa processes with
Higgs radiation off a heavy fermion line in the final state
($e^+e^-\ra f\bar{f}h_i$) on the same footing, we 
discuss the production of $h_1$ in association with heavy fermions:
\begin{equation}
  \label{xsec}
e^+e^- \ra f\bar{f} h_1 \,.
\end{equation}  
Processes (i) and (ii) contribute to this final state when $Z\to
f\anti f$ and $h_2\to f\anti f$, respectively.  If $|C_1|$ is not too
near 1, Eqs.~(\ref{absu},\ref{absd}) imply that radiation processes
(iii) are enhanced when the Higgs boson is radiated off top quarks for
small $\tan\beta$ and off bottom quarks or $\tau$ leptons for large
values of $\tan\beta$.
Since all fermion and Higgs boson masses in the final state must be
kept nonzero, the formulae for the cross section are quite involved.
The tree level expressions can be found in Ref.~\cite{ggk}.  

\begin{figure}[t]
\begin{center}
\hbox to\textwidth{\hss\epsfig{file=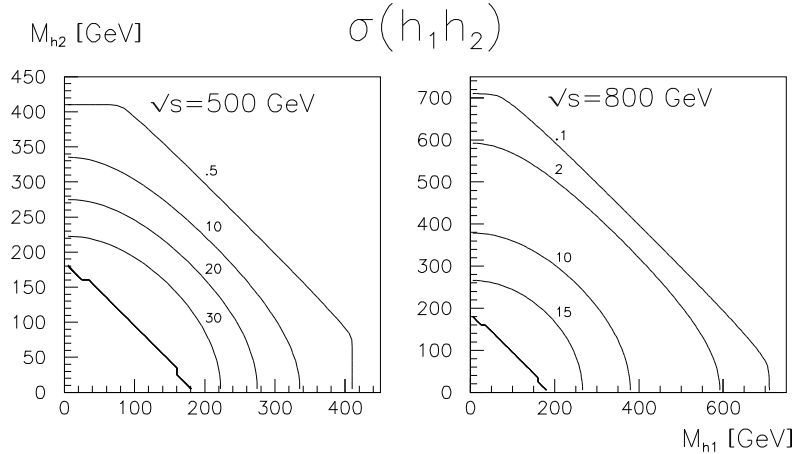,width=12cm,height=12cm}\hss}
\end{center}
\vspace{-6cm}
\caption{Contour lines for ${\rm min} [\sigma(e^+e^-\rightarrow h_1 h_2)]$ 
in fb units, obtained by scanning over the $\alpha_i$
while requiring $\leq 50$ $Zh_1$ or $Zh_2$ events for $L=500\fbi$, 
are plotted in $(m_{h_1},m_{h_2})$ parameter space for the indicated 
$\protect\sqrt{s}$ values and for $\tanb=0.5$. The plots are
virtually unchanged for larger values of $\tanb$. The contour lines 
overlap in the inner corner of each plot as a result of
excluding mass choices inconsistent with experimental 
constraints from LEP2 data.}
\label{h1h2cont}
\end{figure}

Before turning to the case of a single Higgs boson that
is unobservable in Higgs-strahlung or Higgs pair production,
we briefly review and extend to higher energy our earlier results
regarding the detection of a least one of two light Higgs bosons
when $m_{h_1}+m_{h_2},m_{h_1}+\mz,m_{h_2}+\mz<\sqrt{s}$. In
particular, suppose that neither is observable in Higgs-strahlung.
More precisely, as we scan over the 
$\alpha_i$'s we require that the number of  
$\epem \ra Z h_1$ and (separately) the number of $Z h_2$ events  both
be less than 50 for an integrated luminosity of $500 \fbi$.
This will mean that
$|C_1|,\, |C_2|\ll1$, which in turn implies that Higgs-pair
production is at full strength, $|C_{12}|\sim 1$. In Fig.~\ref{h1h2cont},
we show contour plots for the minimum value of the pair production
cross section, min[$\sigma(e^+e^- \rightarrow h_1 h_2)$], as a function of
Higgs boson masses at $\sqrt{s}=$ 500 and 800 GeV obtained by scanning
over mixing angles $\alpha_i$.  With integrated luminosity of $500 - 
1000 \fbi$, a large number of events (large enough to allow for selection
cuts and experimental efficiencies) is predicted for the above
energies over a broad range of Higgs boson masses.  If 50 $h_1h_2$ events
before cuts and efficiencies prove adequate, one can probe reasonably
close to the kinematic boundary defined above. Thus, we will
only need the Yukawa processes for Higgs discovery if (a)
there is only one light Higgs boson or (b) there
are two light Higgs bosons but one cannot
be seen in Higgs-strahlung or Higgs pair production
because the other has full SM strength $ZZ$ coupling.

\begin{figure}[p]
\begin{center}
\hbox to\textwidth{\hss\epsfig{file=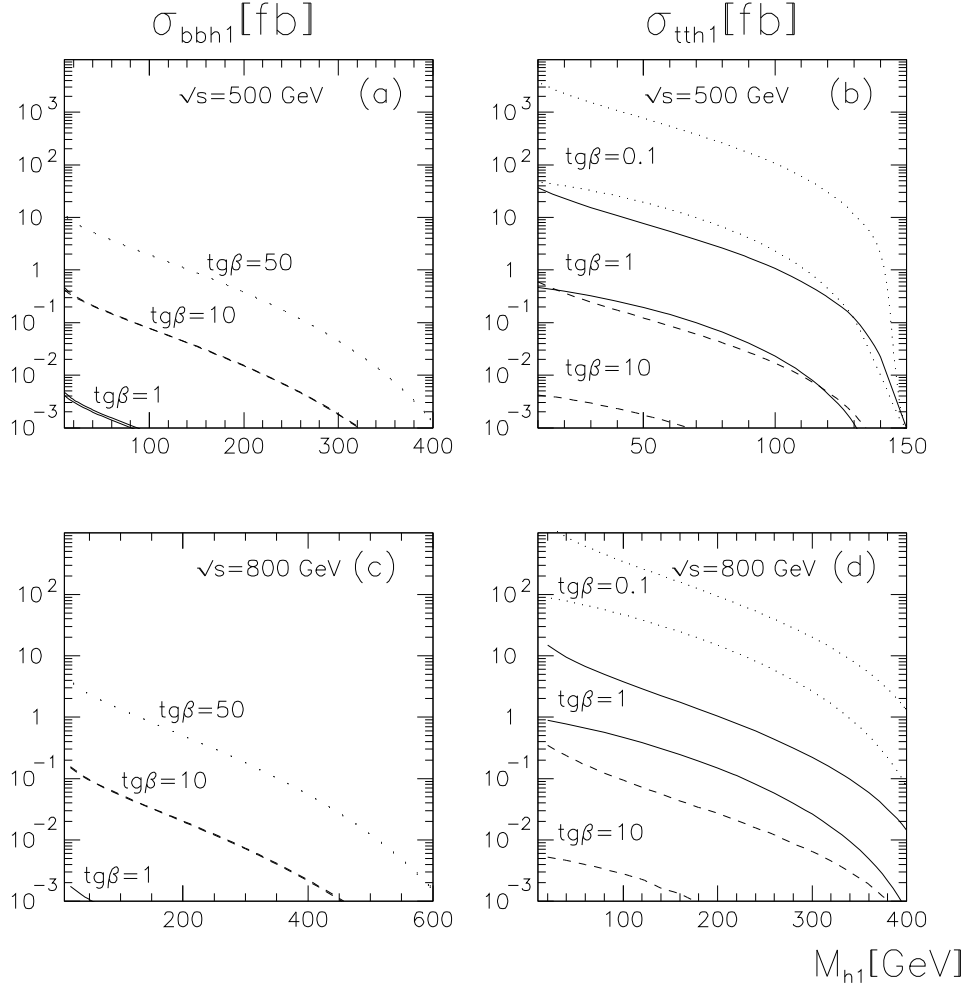,width=14cm,
height=14cm}\hss}
\end{center}
\vspace{-1cm}
\caption{The minimal and maximal values  
of $\sigma(\bbbar h_1)$  and $\sigma( \ttbar h_1)$,
obtained by scanning over $\alpha_1$ and $\alpha_2$ (see footnote 6)
while requiring $\leq 50$ $Zh_1$ events for $L=500\fbi$, are plotted   
for $\protect\sqrt{s}=500$ and 800 GeV. For a given value of $\tanb$,
the same type of line (dots for $\tanb=0.1$ and $t\anti t h_1$, solid
for $\tanb=1$, dashes for $\tanb=10$, dots for $\tanb=50$ and $b\anti bh_1$)
is used for the minimal and maximal values of the cross sections.
In the case of $\bbbar h_1$, the
minimal and maximal values of the cross sections are almost the same.
Masses of the remaining Higgs bosons are assumed to be $1000\gev$.}
\label{yukawalim}
\end{figure}

So now let us turn to the case of a light
Higgs boson that cannot be seen in $Zh_1$ production ($|C_1|\ll 1$)
or Higgs pair production 
($|C_{1i}|\ll 1$, $i=2,3$ and/or $m_{h_2},m_{h_3}>\rts-m_{h_1}$). 
The question is whether the sum rules
(\ref{newsr}) imply that Yukawa couplings are sufficiently large
to allow detection of
the $h_1$ in $t\anti t h_1$ and/or $b\anti bh_1$ production
(assuming both are kinematically allowed). In Fig.~\ref{yukawalim},
we plot the
minimum and maximum values of $\sigma(e^+e^- \rightarrow f\bar{f}h_1)$
for $f=t,b$ as a function of the Higgs boson mass, where we scan over
the mixing angles $\alpha_1$ and  
$\alpha_2$~\footnote{If only the $h_1$ is light, we 
only need to scan over $\alpha_1$ and $\alpha_2$ since all
the couplings of the $h_1$ depend only upon these two mixing angles.}
at a given
$\tanb$ while requiring fewer than $50$ $Zh_1$ events
for $L=500\fbi$.\footnote{We note that, if $C_1\sim 0$, then the minimal
and maximal $b\anti bh_1$ cross sections are almost equal.}
We see that, if $m_{h_1}$ is not large and $\tanb$ is 
either very small or very large, we are guaranteed that
there will be sufficient events in either the $b\bar{b}h_1$ or
the $t\bar{t}h_1$ channel to allow $h_1$ discovery.
However, if $\tanb$ is of moderate size, 
the reach in $m_{h_1}$ is quite limited
if the $\alpha_i$'s are such that $\sigma(t\anti t h_1)$ is minimal.
For example, at $\sqrt{s}=500$ GeV let us take 50 events (before cuts and
efficiencies) as the observability 
criteria.\footnote{For $\tanb\ll 1$ and a light $h_1$, 
requiring 50 $t\anti th_1$ events might not be sufficient
since the $h_1$ will decay predominantly into $c\anti c$, and 
the resulting $t\anti t c\anti c$ final states
will have a large background from  ordinary $t\anti t+$multijet events.
On the other hand, the $t\anti t h_1$ cross section is substantially
enhanced when $\tanb\ll 1$ and, unless there is severe phase
space suppression, we will have substantially more than 50 events.}
For $L=500\fbi$, $\geq 50$ events
then requires $\sigma\geq 0.1\fb$. From Fig.~\ref{yukawalim}, 
we see the following.
\bit
\item At $\tanb=1$, 
$\sigma(b\anti b h_1)\ll 0.1\fb$ for all $m_{h_1}$, while
$\sigma_{\rm min}(t\anti t h_1)$ falls below $0.1\fb$ for
$m_{h_1}>70\gev$. Thus, all but quite light $h_1$'s would elude discovery.
\item
At $\tanb=10$, $\sigma_{\rm min}(t\anti t h_1)\ll
0.1\fb$ and $\sigma_{\rm min}(b\anti b h_1)\simeq \sigma_{\rm
  max}(b\anti b h_1)$ falls below $0.1\fb$ for $m_{h_1}>80\gev$.  
\eit
A $\rts=800$ GeV machine considerably extends the mass reach for $\tanb=1$:
the $h_1$ will be observable for $m_{h_1}\lsim 230\gev$  
(requiring $\sigma_{\rm min}(t\anti t h_1)\geq 0.1\fb$). 
However, for $\tanb=10$,
$\sigma_{\rm min}(t\anti t h_1)$ is again very small 
while $\sigma(b\anti b h_1)$
actually declines faster, falling below $0.1\fb$ already at 
$m_{h_1}\sim 50\gev$. Obviously, for $\tanb$ somewhat less than
10, only a very light $h_1$ is guaranteed to be observable with
only $L=500\fbi$ of integrated luminosity.

For the most part, the minimum cross sections obtained
above when the number of $Zh_1$ events is small and $\tanb$
is moderate in size 
correspond to $\alpha_i$ choices such that $C_1=0$ exactly.
Further, when $C_1=0$, the minimum
cross sections are achieved for a purely CP-odd  $h_1$ 
($\alpha_1=\alpha_2=\pi/2$ and variants thereof), even though
$C_1=-\sb s_1c_2+\cb c_1$ is exactly zero for any choices
of $\alpha_1$ and $\alpha_2$ such that
$c_2=\cotb c_1/s_1$ (see discussion in \cite{ggk}).

For more extreme $\tanb$ values than those
illustrated in Fig.~\ref{yukawalim}, there is, however, an
alternative --- we can actually
zero one of the Yukawa process, namely
the one that is already suppressed, while keeping
the $Zh_1$ cross section small.
For example, if $\tanb$ is large,
implying small $\cb$, $C_1$ can be small enough
to satisfy a finite experimental limit on the number
of $Zh_1$ events if $\alpha_1=0$ ($s_1=0$, $c_1=1$), so that
the first term in $C_1=-\sb s_1c_2+\cb c_1$ is 0. In this extreme, 
the $t\anti t h_1$ cross section will be zero (irrespective of $\alpha_2$)
since $S_1^t\propto -s_1c_2/\sb$ and $P_1^t\propto s_1s_2\cotb$
are both 0. In this limit, $h_1$
is purely $\varphi_1$, the neutral Higgs component
that couples only to bottom quarks.
If $\tanb$ is small ($\sb$ small), the converse situation arises. $C_1$ can be
kept small by taking $c_1=0$ ($\alpha_1=\pi/2$) 
irrespective of the value of $\alpha_2$.
One can then zero the $b\anti b h_1$ cross section
by choosing $\alpha_2=0$.
This is the limit in which $h_1$ is purely $\varphi_2$, the neutral
Higgs component that couples to top quarks.

To illustrate, consider $\rts=800\gev$ and large $\tanb$.
If we require that
$\sigma(Zh_1)<0.1\fb$ (corresponding to fewer than
50 events for $L=500\fbi$), then for $\tanb=10,15,20$
we can choose $\alpha_1=0$, i.e.
$\sigma(t\anti t h_1)=0$, for $m_{h_1}\geq 410,90,0\gev$, respectively.
Note that $\tanb=10$ is just on the border for which this
extreme of zeroing $t\anti t h_1$ becomes relevant.  In fact,
the $\tanb=10$ minimum cross section curve at $\rts=800\gev$
in Fig.~\ref{yukawalim} lies below that which would
be obtained for a purely CP-odd $h_1$ and corresponds to a slight
compromise between exactly zeroing the $t\anti t h_1$ cross
section and the requirement of keeping $\sigma(Zh_1)<0.1\fb$.

\bigskip
\noindent{\bf \boldmath  
Measuring the Yukawa couplings:}
\smallskip

>From the Yukawa sum rules and Fig.~\ref{yukawalim}, it is clear that 
the value of $C_i$ that makes it easiest to measure 
at least one of the $h_i$ Yukawa couplings
is very $\tanb$ dependent.  If $\tanb$ is either $\ll 1$
or $\gg 10$, then $C_i=0$ 
seems to be the most optimistic. This is because
the largest of the minimum cross section values (whether $t\anti t h_i$
for $\tanb\ll 1$ or $b\anti b h_i$ for $\tanb\gg1$) 
is typically substantially enhanced if $|C_i|\sim 0$,
whereas if $|C_i|$ is not small  the sum rules imply
that less enhancement is possible.
In particular, if $|C_i|\sim 1$ (as would be known if the $Zh_i$ rate
is full strength),
then, as outlined earlier, both $\ytsi$ and $\ybsi$ will
be of order unity, approaching 1 exactly if $\tanb$ is either very large
or very small. This implies minimum 
cross section values close to those found for $\tanb=1$. From the $\tanb=1$
minimum cross section curves of Fig.~\ref{yukawalim} for $\rts=500\gev$
($\rts=800\gev$),
one finds that $L=500\fbi$ would not be sufficient for a measurement
of the $b\anti b h_i$ coupling via $b\anti b h_i$ production,
and, if $m_{h_i}$ is
significantly above $70\gev$ ($230\gev$), 
it would also be difficult to measure the
$t\anti t h_i$ Yukawa coupling. 

The situation is quite different if $\tanb$ is moderate in size. 
In this regime of $\tanb$, Fig.~\ref{yukawalim} shows
that a relatively light $h_i$
may not even be observable for $L=500\fbi$ at $\rts=800\gev$ when $|C_i|\ll 1$.
However, it might very well be observable in
one of the Yukawa processes if $|C_i|=1$.  For example,
consider $m_{h_1}=200\gev$ in Fig.~\ref{yukawalim}.  If $\tanb=10$,
$\sigma_{\rm min}(b\anti b h_1)$ and $\sigma_{\rm min}(t\anti t h_1)$
of Fig.~\ref{yukawalim}
are both below $0.1\fb$ if $|C_1|\ll 1$, whereas if $|C_1|=1$
then the $\tanb=1$ curves of Fig.~\ref{yukawalim} become relevant, from which 
we see that $\sigma_{\rm min}(t\anti t h_1)\sim 0.2\fb$, implying that
one could obtain a reasonably good measurement of the $t\anti t h_1$ coupling.

\section{Worst case scenarios}

\begin{figure}[p]
\begin{center}
\hbox to\textwidth{\hss\epsfig{file=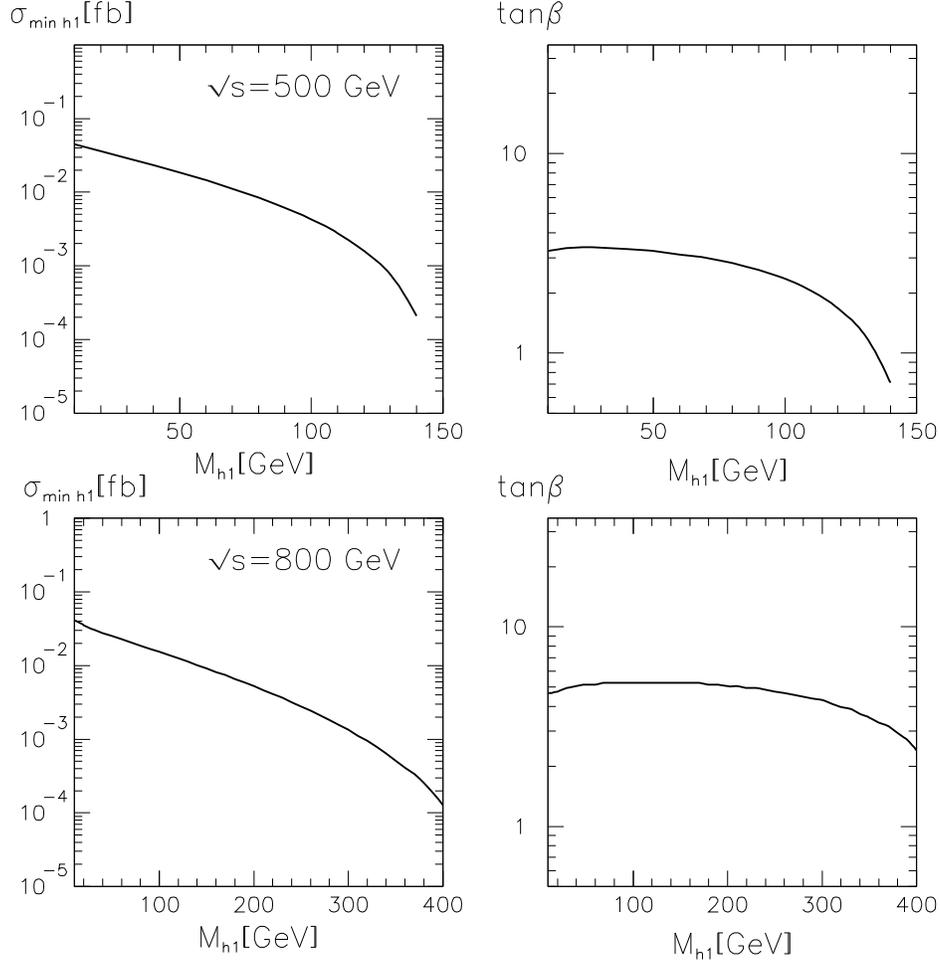,width=14cm,height=14cm}\hss}
\end{center}
\vspace{-1cm}
\caption{
For $\protect\rts=500\gev$ and $\protect\rts=800\gev$, we present as a function
of $m_{h_1}$ the value of $\sigma_{\rm min}\equiv {\rm min}_{\tanb}\left(
{\rm min}_{(\alpha_1,\alpha_2)}\left\{{\rm max}[
\sigma_{\rm min}(b\anti b h_1),
\sigma_{\rm min}(t\anti t h_1)]\right\}\right)$, 
and the corresponding value of $\tanb$, as obtained by scanning
over $\tanb$ and $(\alpha_1,\alpha_2)$ parameter space
subject to constraints (I) and (II) --- see text for details.
Masses of the remaining Higgs bosons are assumed to be 1000 GeV.}
\label{worstcase}
\end{figure}

The discussion of the previous section raises the
interesting question of just how much luminosity
is required as a function of $\rts$ and $m_{h_1}$
in order to absolutely guarantee discovery of a light $h_1$
in at least one of the three modes, 
$Zh_1$, $t\anti t h_1$ or $b\anti b h_1$. 
We consider only ${h_1}$ masses such that both Yukawa modes are kinematically
allowed. In our first plot, Fig.~\ref{worstcase}, we impose
the requirements (I) that there be $\leq 50$ $Zh_1$ events for $L=500\fbi$
and (II) that LEP/LEPII upper bounds \cite{opal_limit}
on the $ZZh_1$ coupling be satisfied. 
For each $m_{h_1}$, we scan over $\tanb$, to determine 
the $\tanb$ at which ${\rm min}_{(\alpha_1,\alpha_2)}\,
\left\{{\rm max}\left[\sigma(b\anti b h_1),
\sigma(t\anti t h_1)\right]\right\}$ 
is smallest. Here, ${\rm max}\left[\sigma(b\anti b h_1),
\sigma(t\anti t h_1)\right]$ is the larger of $\sigma(b\anti bh_1)$
and $\sigma(t\anti th_1)$ for any given $(\alpha_1,\alpha_2)$ choice,
and ${\rm min}_{(\alpha_1,\alpha_2)}$
refers to the minimum value of this maximum after
scanning over all $(\alpha_1,\alpha_2)$ values (see footnote 6)
satisfying the constraints (I) and (II). 
We then look for
the $\tanb$ value at which this minimum is smallest.
For any other $\tanb$ choice, one or the other cross section will
be larger than this minimum for {\it all} 
choices of $(\alpha_1,\alpha_2)$ and the corresponding mode easier
to observe. This defines the `worst case' $\tanb$ choice
for which a light Higgs boson that is unobservable in the $Zh_1$ mode
will be most difficult to see by virtue of neither the
$b\anti b h_1$ nor the $t\anti t h_1$ cross section
being enhanced relative to the other. 
In Fig.~\ref{worstcase},
we plot the worst case choice of $\tanb$ and the corresponding
value of 
 $\sigma_{\rm min}\equiv {\rm min}_{\tanb}\left({\rm min}_{(\alpha_1,\alpha_2)}
\left\{{\rm max}\left[\sigma(b\anti b h_1),
\sigma(t\anti t h_1)\right]\right\}\right)$.
Results are presented for both $\rts=500\gev$ and $\rts=800\gev$.

We observe that the integrated luminosity required for the worst
case cross section to yield 50 events 
in each of the Yukawa modes is always greater than $L=500\fbi$.
Even for small $m_{h_1}\sim 10\gev$, 
$\sigma_{\rm min}\sim 4-5\times 10^{-2}\fb$ at these energies,
implying that $L\gsim 1000\fbi$ would be required for just $40-50$ events
in each. As we increase $m_{h_1}$, the worst case cross section
for $\rts=500\gev$ falls dramatically and detection
of the $h_1$ would not be possible for any reasonable $L$.
However, at $\rts=800\gev$ the worst case cross section
has only fallen to about 
$1.7\times 10^{-2}\fb$ at $m_{h_1}=100\gev$ 
for which $L=3000\fbi$ would yield about 50 events in the $b\anti b h_1$
and $t\anti t h_1$ modes, each (while still not guaranteeing
as many as 50 $Zh_1$ events). Possibly, such a large $L$
could be achieved after several years of running.   

\begin{figure}[h]
\begin{center}
\hbox to\textwidth{\hss\epsfig{file=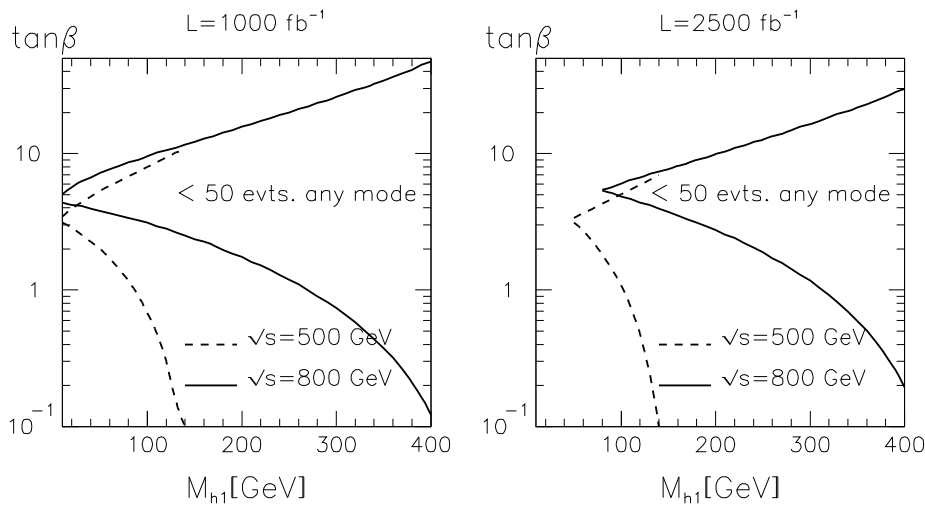,width=14cm,height=14cm}\hss}
\end{center}
\vspace{-7.5cm}
\caption{
For $\protect\rts=500\gev$ (dashes) and $\protect\rts=800\gev$ (solid)
we present as a function of $m_{h_1}$ the maximum and minimum $\tanb$
values between which $t\anti t h_1$, $b\anti b h_1$ and $Zh_1$
final states can (for some choice of $(\alpha_1,\alpha_2)$
consistent with constraint (II) --- see text)
all have fewer than 50 events assuming
(a) $L=1000\fbi$ or (b) $L=2500\fbi$.
Masses of the remaining Higgs bosons are assumed to be 1000 GeV.}
\label{worstcaseband}
\end{figure}

To illustrate all of this more completely,
we have determined, as a function of $m_{h_1}$, the $\tanb$ range for which 
constraint (II) above is satisfied while the $Zh_1$, the
$b\anti b h_1$ and the $t\anti t h_1$ cross section {\it each} yield
fewer than 50 events for at least one $(\alpha_1,\alpha_2)$
choice assuming (a) $L\leq 1000\fbi$ or (b) $L\leq 2500\fbi$ 
and $\rts=500\gev$ or, separately, 
$\rts=800\gev$. These $\tanb$ ranges are represented in 
Fig.~\ref{worstcaseband} by the wedge of $\tanb$
between the solid ($\rts=800\gev$) or dashed ($\rts=500\gev$) lines.
For $\tanb$ values above (below) the upper (lower) line,
$b\anti b h_1$ ($t\anti t h_1$) will be observable
for all $(\alpha_1,\alpha_2)$ choices. We see that, even after
combining $\rts=500\gev$ and $\rts=800\gev$ running, the $L=1000\fbi$ 
wedge begins at 
$m_{h_1}\sim 25\gev$ and widens rapidly with increasing $m_{h_1}$.
For $L=2500\fbi$, the wedge begins at a higher $m_{h_1}$ value ($\sim 80\gev$
for $\rts=800\gev$), but still expands rapidly as $m_{h_1}$ increases further.
Thus, it is apparent that, despite the sum rules guaranteeing significant
fermionic couplings for a light 2HDM Higgs boson that is unobservable
in $Z$+Higgs production, $\tanb$ and the $\alpha_i$ mixing angles
can be chosen so that the cross
section magnitudes of the two Yukawa processes are simultaneously
so small that detection of such an $h_1$ cannot be guaranteed for integrated
luminosities that are expected to be available.

On a final technical note, we have found that the
$h_1$ is, for the most part, either exactly, or almost exactly, CP-odd 
for the $(\alpha_1,\alpha_2)$ parameters corresponding to 
the curves plotted in Figs.~\ref{worstcase} and \ref{worstcaseband}.
The only exception is for $m_{h_1}$ values between $\sim 160\gev$ 
($\sim240\gev$) and
$\sim 270\gev$ ($\sim 300\gev$) for $L=1000\fbi$
($L=2500\fbi$) at $\rts=800\gev$ along
the upper lines in Fig.~\ref{worstcaseband}.
For this range, $\alpha_1$ can be chosen close to 0 to minimize 
$\sigma(t\anti t h_1)$ while continuing to satisfy the $Zh_1$ event
number constraint (see the discussion at the end of the previous section).

\section{Discussion and conclusions}

The sum rules, Eq.~(\ref{newsr}),
relating the Yukawa and Higgs-$ZZ$ couplings of a general CP-violating
two-Higgs-doublet model have important
implications for Higgs boson discovery at an $\epem$
collider. In particular, for any $h_i$, if the $ZZh_i$
coupling is small, then the $t\anti t h_i$
or $b\anti b h_i$ Yukawa coupling must be substantial. 
This means that any one of the three neutral
Higgs bosons that is light enough to be produced in $\epem\to t\anti t h_i$
(implying that $\epem\to Zh_i$ and 
$\epem\to b\anti b h_i$ are also kinematically allowed) will normally
be found at an $e^+e^-$ linear collider if the
integrated luminosity is sufficient.
However, we have found that the mass reach in $m_{h_i}$ may
fall well short of the $\rts-2\mt$ kinematic limit for 
moderate $\tanb$ values and anticipated luminosities.  
We have made a precise determination
of the value of $\tanb$ (as a function of $m_{h_i}$) for which
the smallest (common) value of the $t\anti t h_i$ and $b\anti b h_i$
cross sections is attained
when the $ZZh_i$ coupling is suppressed.  From this, we have computed
as a function of $m_{h_i}$ the minimum luminosity required
in order to detect such an $h_i$. Even at $\rts=800\gev$,
to guarantee detection of a Higgs boson with small $ZZ$
coupling for the worst possible choice of $\tanb$ and neutral
Higgs sector mixing angles
would require an integrated luminosity in excess
of $1000\fbi$ starting at $m_{h_i}\sim 10\gev$. Further, the minimum $L$
required to guarantee detection for the worst choices
of $\tanb$ and mixing angles increases rapidly as $m_{h_i}$ increases,
as does the band of $\tanb$ in which $L>1000\fbi$ is required.

We also discussed the case of an $h$ that {\it is}
 observed in the $Zh$ final state
but also light enough to be seen in $t\anti t h$ and, by implication,
$b\anti b h$. 
We have noted that if $Zh$ production proceeds with SM strength,
then the same sum rules can be used to show that measurement of
its  $b\anti b$ coupling will
be impossible for any conceivably achievable integrated
luminosity, while measurement of its $t\anti t$ coupling
may only be possible for $m_h$ up to values
significantly below the $\rts -2\mt$ phase space limit
(the exact reach depending upon the integrated luminosity
and Higgs sector mixing angles).


Finally, we note that detection  `guarantees' for the 2HDM 
model are likely to apply over an even more restricted
range of model parameter space in the case of
the Tevatron and LHC hadron
colliders.  In the case of the Tevatron, the small rate for $t\anti
t$+Higgs production is a clear problem. In the case of the LHC, a
detailed study is needed to determine what
cross section level is required in order that Higgs
detection in the $t\anti t$+Higgs 
channel will be possible.
Existing studies in the context of supersymmetric models can
be used to point to parameter regions that are problematical because
of large backgrounds and/or signal dilution due to sharing of
available coupling strength.  Almost certainly,
very small (large) $\tanb$ values will be needed
in order to be certain that the $t\anti t$+Higgs ($b\anti b$+Higgs)
modes will be viable. Still, it is clear that the sum rules
do imply that difficult parameter regions are of limited extent.

\vspace{1.5cm}
\centerline{\bf Acknowledgments}
\vspace{.5cm} This work was supported in part by the Committee for
Scientific Research (Poland) under grants No. 2~P03B~014~14, No.
2~P03B~030~14, by Maria Sklodowska-Curie Joint Fund II
(Poland-USA) under grant No. MEN/NSF-96-252, by the U.S.
Department of Energy under grant No. DE-FG03-91ER40674
and by the U.C. Davis Institute for High Energy Physics.

\end{document}